\documentstyle{l-aa}

\begin{document}

\def\bi{{B\,{\sc i}}}
\def\multi{{\sc multi}}
\def\lte{{\sc lte}}
\def\ee#1{\cdot 10^{#1}}
\def\feh{{[Fe\H]}}
\def\doepsf{1}       
\ifnum\doepsf=1\input{epsf}

\thesaurus{02.12.1, 08.01.1, 08.01.3, 10.01.1}

\title{ The NLTE formation of neutral-boron lines in cool stars}
\author{Dan Kiselman \inst{1,2} \and Mats Carlsson \inst{3}}
\institute{NORDITA, Blegdamsvej 17, DK-2100 Copenhagen, Denmark \and
The Royal Swedish Academy of Sciences, Stockholm Observatory,
S-133 36 Saltsj{\"o}baden, Sweden \and
Institute of Theoretical Astrophysics, University of Oslo, Box 1029,
Blindern, N-0315 Oslo, Norway}

\offprints{Dan Kiselman (Stockholm Observatory)}
\date{Received 4 October 1995 / Accepted 21 December 1995}
\maketitle


\begin{abstract}
We study the formation of \bi\ lines in a grid of cool stellar
model atmospheres without the assumption of local
thermodynamic equilibrium (\lte). The non-\lte\ modelling
includes the effect of other lines blending with the \bi\ resonance lines.
Except for the cases where the \bi\ lines are
very strong, the departures from \lte\ relevant
for the resonance lines can be described as
an overionisation effect and an optical-pumping effect.
This causes the lines to be weaker than in \lte\ so that
an abundance analysis assuming \lte\ will underestimate stellar
boron abundances. 
We present non-\lte\ abundance corrections useful to improve on abundances
derived from the \bi\,250\,nm and 209\,nm lines under the \lte\ assumption.
Application of the results on literature data indicates
that the B/Fe ratio in metal-poor stars is constant.

\smallskip

\end{abstract}

\section{Introduction}
The Hubble Space Telescope has made it possible to derive boron
abundances for cool stars by observation of the \bi\ resonance
doublet at 250\,nm. (Duncan et al. 1992; Lemke et al. 1993; 
Edvardsson et al. 1994; Duncan et al. 1995; more publications will certainly
follow.) These abundances,
for stars of different ages and in different evolutionary stages,
can give important clues to the origin of the light elements and 
to stellar evolution.
It is important that the abundances derived are correct, so 
the validity of the stellar abundance analyst's standard
approximations of plane-parallel stellar atmospheres with
line formation in local thermodynamic equilibrium (\lte)
must be investigated.

Kiselman (1994) investigated the statistical equilibrium of 
neutral boron in three solar-type atmospheric models and found
significant departures from \lte\ in stars
hotter or more metal-poor than the Sun.
The most important process causing these departures was found to be the
pumping in the ultraviolet resonance lines that gives rise to
line source functions exceeding the local
Planckian value and to overionisation.
The effects on derived abundances
are significant when it comes to interpretation of observations
of metal-poor halo stars (Edvardsson et al. 1994).

The aim of this paper is to investigate the formation of \bi\ lines
in cool stars and to provide non-\lte\ abundance
corrections in a form that is convenient to use for general
observers. To this end we have studied the \bi\ lines in a
grid of model atmospheres and for a range of boron abundances.
We will first present the techniques used, then discuss the
line-formation circumstances and their dependence on atmospheric
parameters, and finally present abundance corrections that
are of practical use for abundance analysts using \lte\
spectral synthesis.
The discussion concentrates on the lines which are of 
observational interest,
the least blended in each of the 250\,nm and 209\,nm doublets,
that is
$\lambda_{\rm vac} = 249.75\,{\rm nm}$ and $\lambda_{\rm vac} =
209.02\,{\rm nm}$
($\lambda_{\rm air} = 249.68\,{\rm nm}$ and $\lambda_{\rm air} =
208.96\,{\rm nm}$), all wavelengths from Johansson et al. 1993).
These lines will be referred to as the 250\,nm and the 209\,nm lines.
The 209\,nm doublet holds some promise in being useful for determining
boron isotopic ratios (Johansson et al. 1993), but no detailed
observational analysis of these lines have yet been published.

\begin{figure*} 
\ifnum\doepsf=1\hspace*{.2cm}\epsfxsize=16cm\epsffile{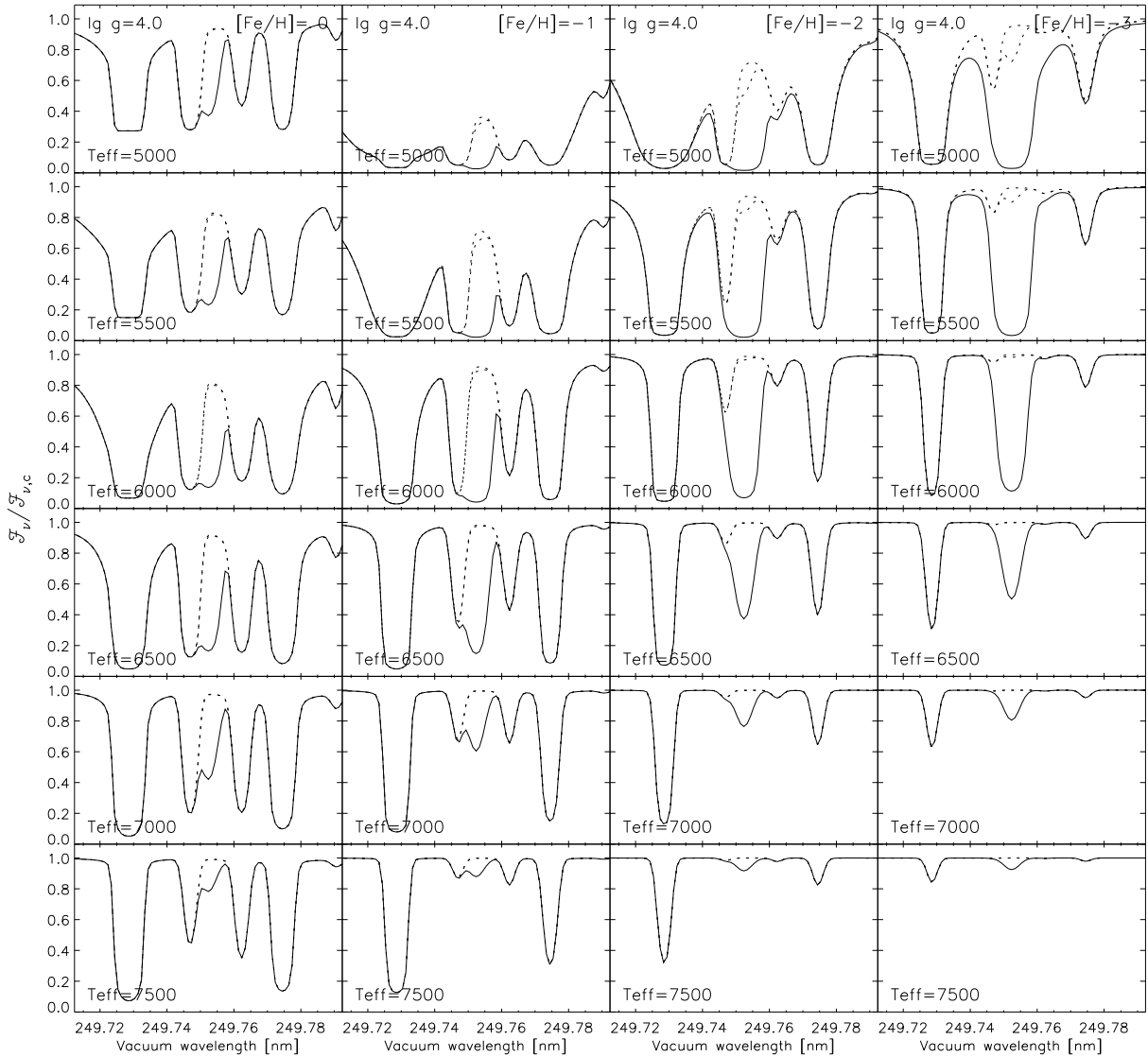}\else\pic
place{1cm}\fi
\caption[]{Synthetic flux spectra for the region around
\bi\,249.75\,nm, the section of the grid with $\lg g = 4.0$. 
The spectra have been normalised to the continuum level as computed
without lines. 
No instrumental, macro-turbulent, or rotational
broadening have been applied. Solid curve: $A_{\rm B} = 2.6$. 
Dotted curves: $A_{\rm B} = 0.0$ and no boron}
\label{fig_bspec_250}
\end{figure*}

\begin{figure*}
\ifnum\doepsf=1\hspace*{.2cm}\epsfxsize=16cm\epsffile{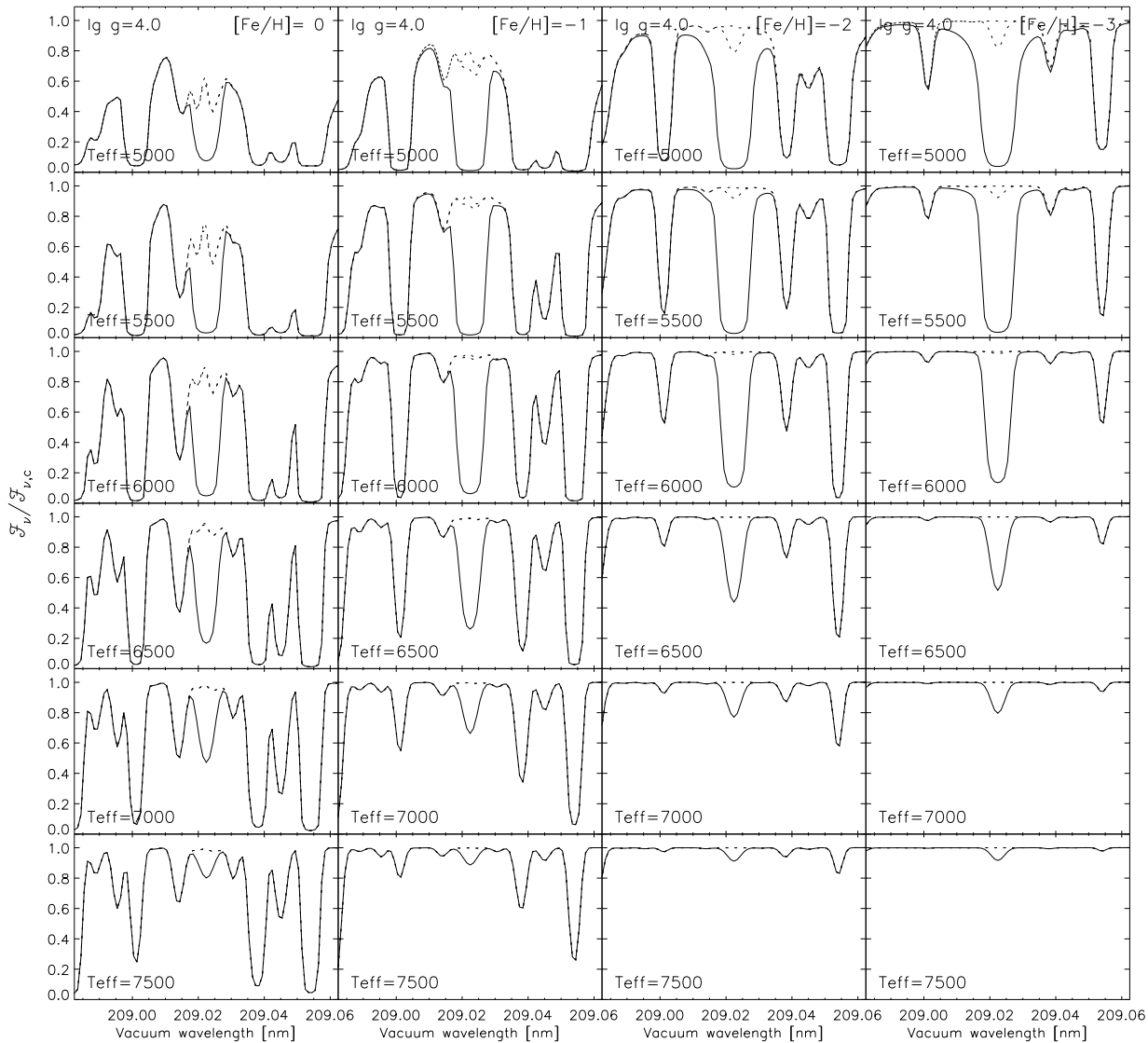}\else\pic
place{1cm}\fi
\caption[]{Synthetic flux spectra for the region around
\bi\,209\,nm, same treatment as in Fig.~\ref{fig_bspec_250}. Solid curve: $A_{\rm B} = 2.6$. 
Dotted curves: $A_{\rm B} = 0.0$ and no boron}
\label{fig_bspec_209}
\end{figure*}

\section{Methods}
\subsection{Conventions}
We express elemental abundances on the customary scale
where the abundance $$A_{\rm X} = \lg N({\rm X})/N({\rm H}) +12,$$
with $\lg$ being the logarithm, base 10.
 
Overall metallicity is given relative to solar
as $$[{\rm Fe}/{\rm H}] = \lg {N({\rm Fe}) \over N({\rm H})} -
\lg {N({\rm Fe}) \over N({\rm H})}_{\sun}.$$
To comply with common usage in stellar atmosphere work, 
the logarithm of the surface gravity, $\lg g$ is given
with $g$ in cgs units.

Quantities denoted with an asterisk, e.g. $N^*$, are \lte\ values.

\subsection{Model atmospheres}
The model atmospheres were produced using the code {\sc osmarcs}
of Edvardsson et
al. (1993) which is an extended version of the {\sc marcs} program
(Gustafsson et al. 1975). This code includes a great number of
spectral lines in the blue and ultraviolet by means of opacity
sampling. The line data are from the Kurucz compilation 
(e.g. Kurucz 1991) and
it has been claimed (Kurucz 1991; Edvardsson et al. 1993) that 
inclusion of these opacities solves the so-called ``missing-opacity
problem'' for the blue and ultraviolet parts of the solar spectrum. 
It has, however, been questioned (Bell et al. 1994, see also Gustafsson 1995
and Bell \& Tripicco 1995)
whether the adding of these line opacities really solve the problem 
in the right way,
i.e. the resulting models may reproduce the observed flux spectrum 
well at low resolution but badly at high resolution. 
If this is the case, systematic errors
may appear in comparisons of different stars.
We consider, however, these models as the best choice,
since they apparently are successful in reproducing the ultraviolet fluxes
of solar-type stars of different metallicities 
(Edvardsson et al. 1993, Duncan et al. 1992), something which is
crucial for the current study.

The radiation-field values for different depths and wavelengths
that are produced by the atmosphere code are saved and
used in the non-\lte\ spectral modelling for the calculation of
the fixed radiative rates.

The parameters of the atmospheric grid are displayed in 
Table~\ref{tab_grid_params}.
The temperature and gravity ranges were determined
by the limitations of the {\sc osmarcs} code.
We are aware that some of the combinations of fundamental parameters
are unlikely to correspond to any observed stars. 
All models were calculated with a Gaussian microturbulent velocity
of 2.0\,km\,s$^{-1}$, this was also used in the subsequent line calculations.

The relative elemental abundances assumed for the atmospheres and for
the background opacities in the non-\lte\ calculations are 
the solar abundances of Anders \& Grevesse (1989). The
exception is iron, for which we use the meteoritic abundance
of 7.51 which is close to what analyses of Fe\,{\sc ii} lines
give (e.g. Holweger et al. 1990, Bi\'emont et al. 1991).

\begin{table}
\caption[]{Fundamental parameters of the atmospheric grid}
\label{tab_grid_params}
\begin{flushleft}
\begin{tabular}{ll}
\hline\noalign{\smallskip}
$T_{\rm eff}$\,[K] &$=$ 5000, 5500, 6000, 6500, 7000, 7500\\
\noalign{\smallskip}
$\lg g$\,[cgs]     &$=$ 2.00, 3.50, 4.00, 4.50\\
\noalign{\smallskip}
[Fe/H]             &$=$ 0.00, $-$1.00, $-$2.00, $-$3.00\\
\noalign{\smallskip}
$\xi_{\rm t}$\,[km\,s$^{-1}$]      &$=$ 2.00\\
\noalign{\smallskip}
\hline
\end{tabular}
\end{flushleft}
\end{table}

\subsection{Model atom}
Our \bi\ model atom is the one compiled by 
Kiselman (1994) with minor refinements.
It has 30 bound levels and one continuum level and features 
114 line transitions.
The 30 photoionisation transitions are treated as fixed rates,
calculated from the radiation field data from the model atmospheres.
Photoionisation cross sections, $f$-values, collisional cross sections
for transitions between the lower levels are theoretical data
of considerable accuracy. For the lines of 
observational interest, we use the laboratory $f$-values 
of Johansson et al. (1993).

\subsection{Non-LTE code}
We use the operator perturbation technique of Scharmer \& Carlsson (1985)
as coded in the program \multi\ by Carlsson (1986).
We employ \multi\ version 2.2, in which line blanketing is taken
into account in the photoionisation, the treatment of background
opacities is improved and the many-level treatment is speeded up by using
the local operator of Olson et al. (1986).
This version also allows
the inclusion of spectral lines in the background opacities for
the line transitions treated in detail, as
described in the following section.

\subsection{The problem of background lines}
Kiselman (1994) did not include the effect of other lines blending
with the \bi\ resonance lines -- the ``background'' opacities were
just the continuous ones. The effect of such blending can
be expected to be a moderation of the pumping effect that causes
the departures from \lte. This will tend
to decrease the positive non-\lte\ abundance corrections.

A full non-\lte\ treatment of blended lines of different atomic
species
is however a formidable task. In our calculations, the ``background''
lines are assumed to be formed in \lte\ and in pure absorption.
(See, however, our discussion in section~\ref{sec_back_lines}.)

It is currently not possible to assemble detailed line lists for the
ultraviolet 
spectral regions of all the \bi\ resonance lines. Our treatment of blending
is schematic for the shortest-wavelength lines but
gets more detailed towards longer wavelengths:

\begin{itemize}
\item The shortest-wavelength resonance lines are treated in the same
way as the photoionisations, i.e. the transitions are considered to be
fixed, with radiation fields computed by the atmospheric code, thus
including the effect of line blanketing in an averaged way. 

\item For the doublet at 185\,nm, background lines from the Kurucz data
set are introduced, but without any observational checks on the result.

\item For the region of the 209\,nm doublet,
a preliminary line list was kindly provided
by Dr. B. Edvardsson. The resulting synthetic solar spectrum looked
satisfying when compared 
to an uncalibrated spectrum from the Solar Maximum
Mission satellite that was kindly made available by P{\aa}l Brekke.

\item The background lines for the 250\,nm doublet are those used in
the spectrum synthesis of Duncan et al. (1992). 
Some new lines from the Kurucz data set were introduced
to approximately fit the Procyon spectrum displayed by  
Lemke et al. (1993). (See the spectral region between 2498.0\,\AA\ and
2498.3\,\AA\ (air) in their Fig. 2.) The Fe\,{\sc ii} line at
2497.820\,{\AA} (air) was removed from the Duncan et al. list.

\end{itemize}

The calculation of the Boltzmann-Saha populations needed for the
background line opacities is handled by the continuous-opacity
package. This means that certain simplifications have to be made, for
example, some elements have to be treated as others of similar
ionisation potential: Ti and V as Cr, Mn and Co as Fe.

Note that it is not our aim
to produce detailed synthetic spectra that can be directly compared
with observations, but rather to study the departures from \lte\ and
to acquire differential 
abundance corrections. The resulting spectra are, however,
illustrative, and some are displayed in Figs.~\ref{fig_bspec_250} and \ref{fig_bspec_209}.

\subsection{The grid}
The results presented here come from running the program 
 for the atmospheric grid and 21 boron
abundances
($-$1.5, $-$1.2, $-$0.9, ..., 4.2, 4.5) plus effectively zero
boron abundance.
For diagnostic purposes, runs with boron abundances 0.0 and 2.6 were
made, for which the full computation results were saved for
the lines of observational interest, including
radiation-field data.
All calculations were made with and without the background
lines. The results from both these data sets are discussed in the
following. For the sake of understanding the line-formation
circumstances, we will first look at the results without
background lines and view the adding of the lines as
a perturbation of this case. But it is the results
from the computations with background lines that
will be used for the final recommended abundance corrections.

\begin{figure} 
\ifnum\doepsf=1\hspace*{.2cm}\epsfxsize=8.8cm\epsffile{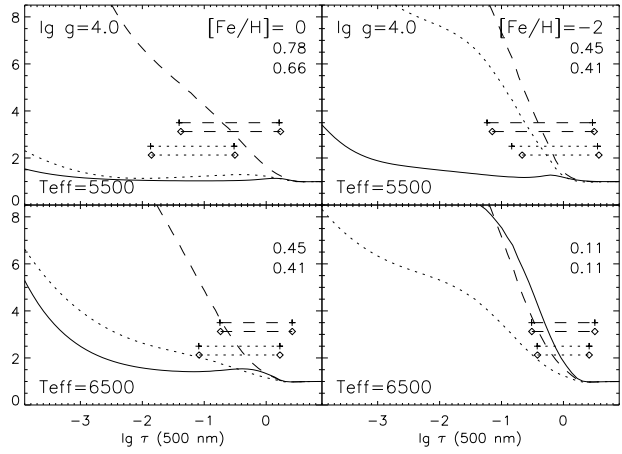}\else
\picplace{1cm}\fi
\caption[]{Departure coefficients, $A_{\rm B} = 0.0$, no background lines. 
Solid curve: depletion coefficient of neutral boron
$d_0 = N^*({\rm \bi})/N({\rm \bi})$. Dotted curve: $S^l/B_\nu$ for
\bi\,250\,nm. Dashed curve: $S^l/B_\nu$ for \bi\,209\,nm. Crosses
show where the total optical depth in the line centre equals 1/3 and 3
(styles of connecting lines identify spectral lines as before),
diamonds show the same thing for the continuum. Floating-point numbers
give the departures from \lte\ in equivalent widths:
$W/W^*$ for 250\,nm (upper numbers) and for 209\,nm (lower numbers) }
\label{fig_depcoff_n0g4}
\end{figure}
\begin{figure} 
\ifnum\doepsf=1\hspace*{.2cm}\epsfxsize=8.8cm\epsffile{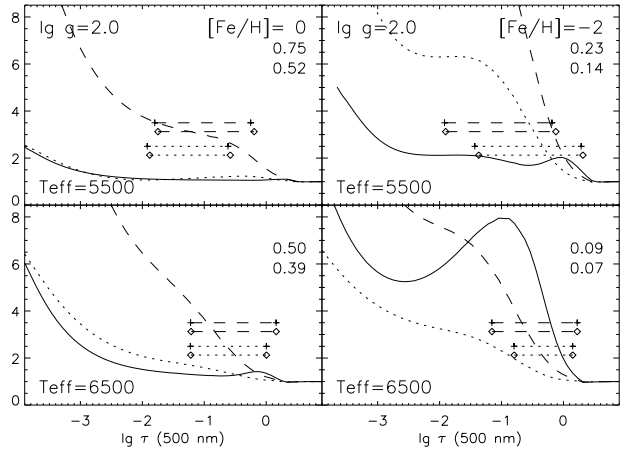}\else\picplace{1cm}\fi
\caption[]{As Fig. \ref{fig_depcoff_n0g4}, for lower-gravity atmospheres}
\label{fig_depcoff_n0g2}
\end{figure}
\begin{figure} 
\ifnum\doepsf=1\hspace*{.2cm}\epsfxsize=8.8cm\epsffile{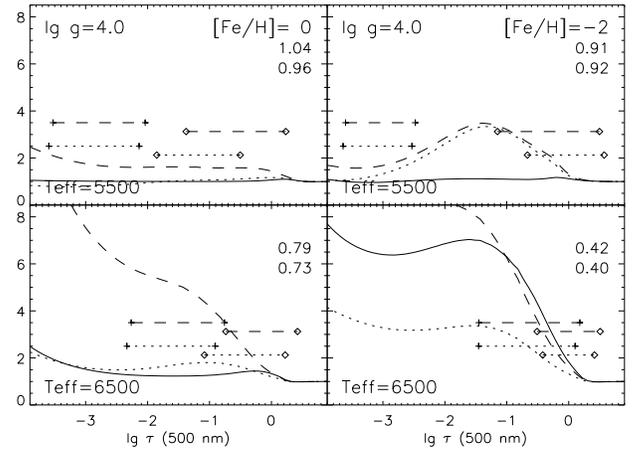}\else\picplace{1cm}\fi
\caption[]{Departure coefficients, $A_{\rm B} =
2.6$. Designations and symbols as in Fig. \ref{fig_depcoff_n0g4}}
\label{fig_depcoff_n26g4}
\end{figure}
\begin{figure} 
\ifnum\doepsf=1\hspace*{.2cm}\epsfxsize=8.8cm\epsffile{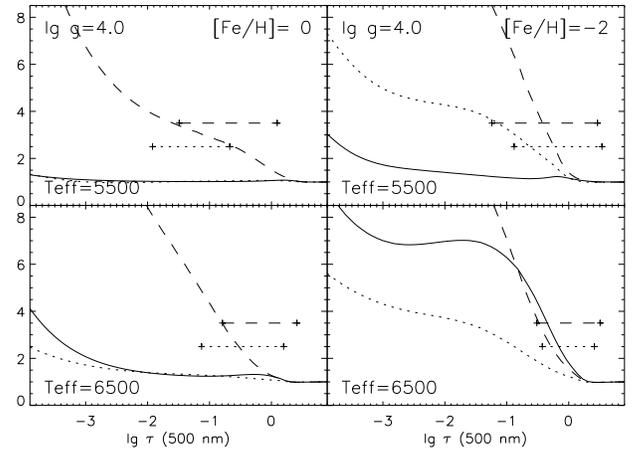}\else\picplace{1cm}\fi
\caption[]{Departure coefficients, $A_{\rm B} = 0.0$, with 
background lines. Designations and symbols as in Fig. \ref{fig_depcoff_n0g4}}
\label{fig_depcoff_a0g4}
\end{figure}

\begin{figure} 
\ifnum\doepsf=1\hspace*{.2cm}\epsfxsize=8.8cm\epsffile{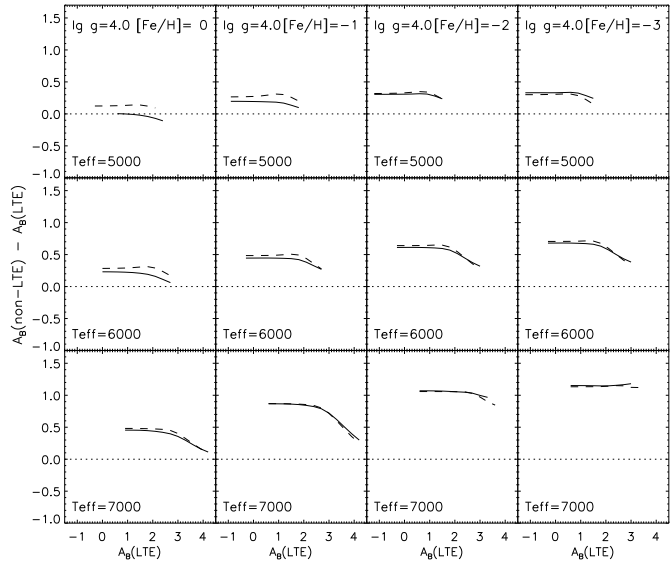}\else\picplace{1cm}\fi
\caption[]{Non-\lte\ abundance corrections, no background lines.
Solid curve: \bi\,250\,nm. Dashed curve: \bi\,209\,nm
}
\label{fig_abcorr_ng4}
\end{figure}

\begin{figure}
\ifnum\doepsf=1\hspace*{.2cm}\epsfxsize=8.8cm\epsffile{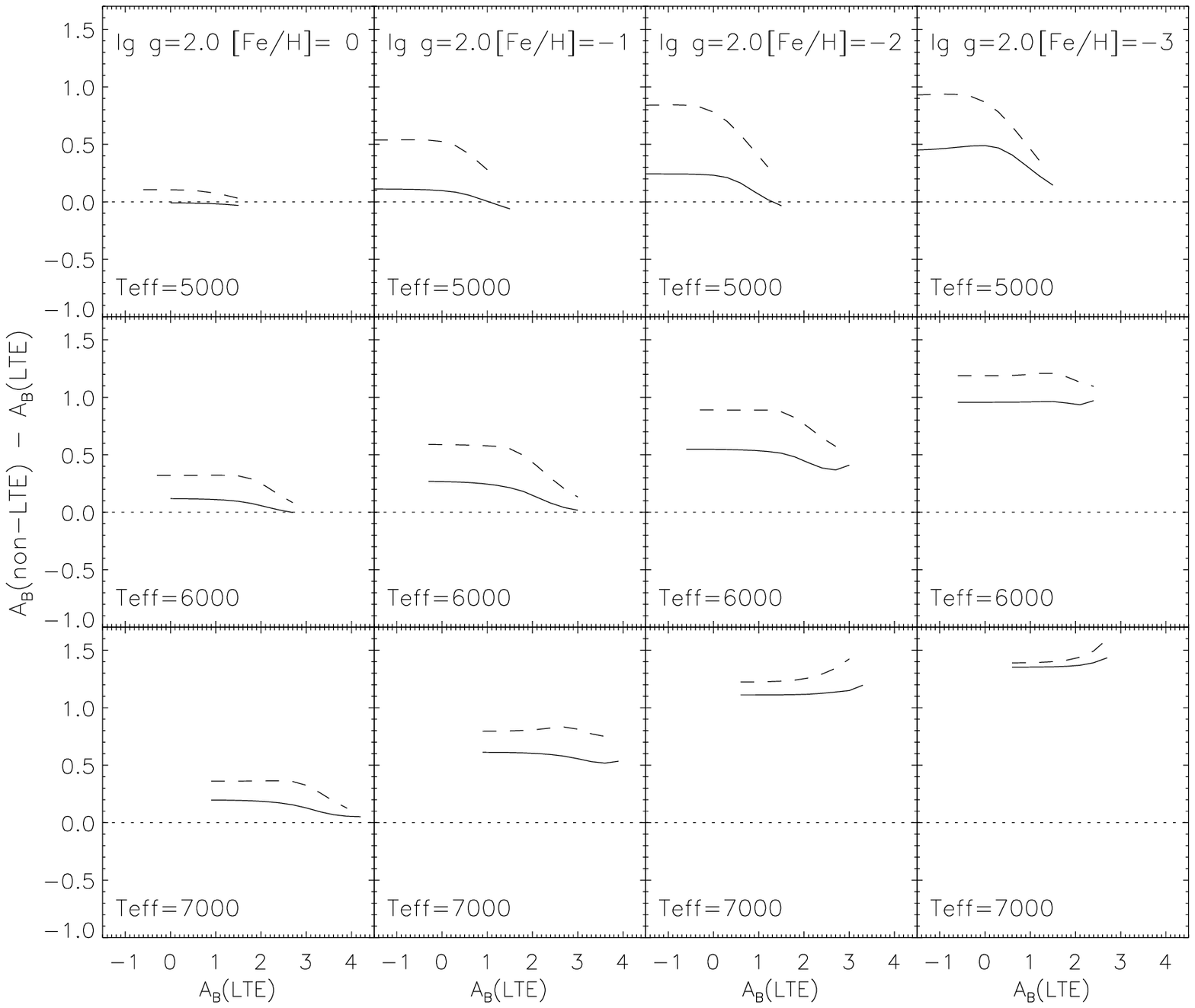}\else\picplace{1cm}\fi
\caption[]{Non-\lte\ abundance corrections, lines included
in background opacities. Solid curve: \bi\,250\,nm. Dashed curve: \bi\,209\,nm
}
 \label{fig_abcorr_ag2}
\end{figure}

\begin{figure} 
\ifnum\doepsf=1\hspace*{.2cm}\epsfxsize=8.8cm\epsffile{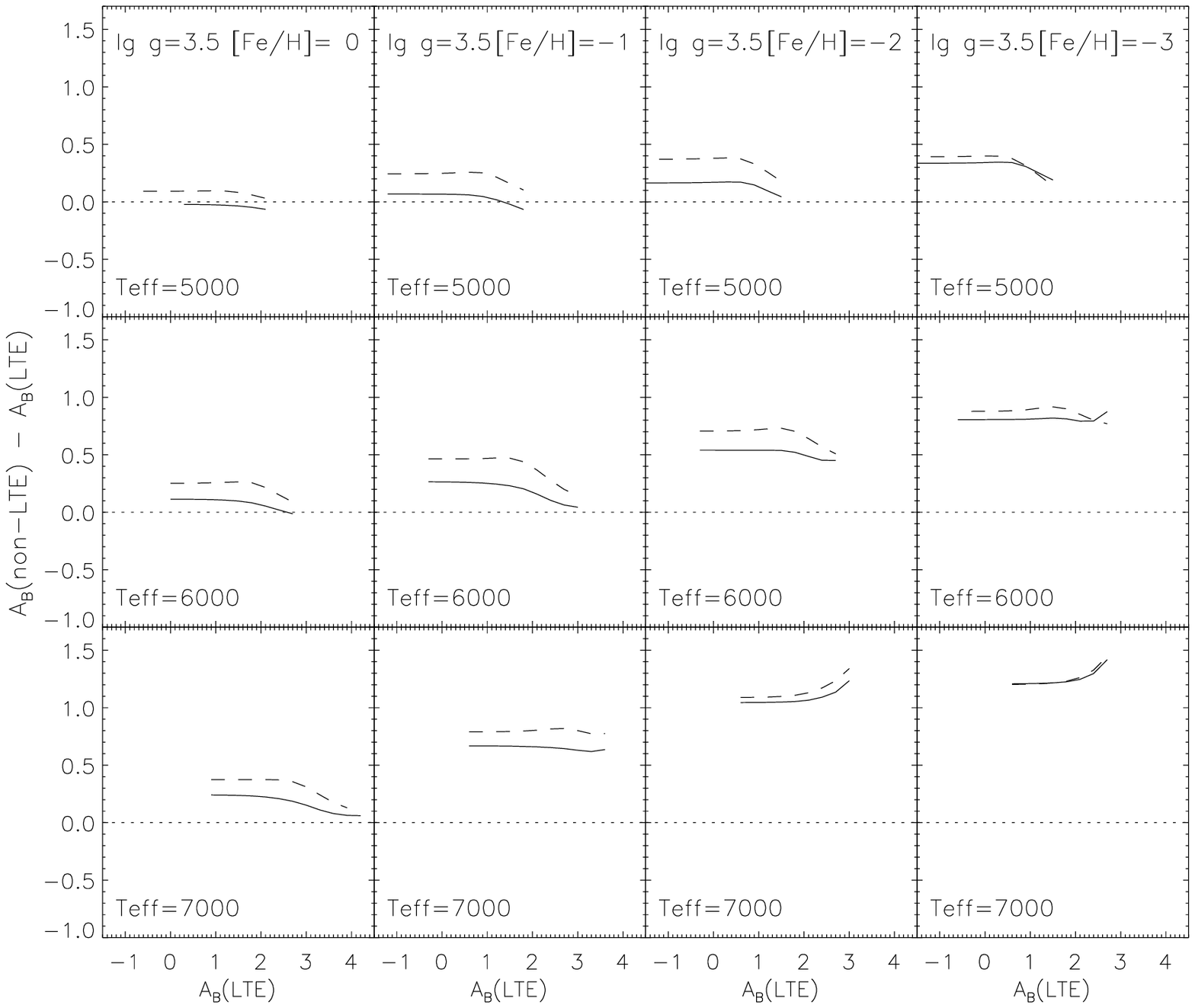}\else\picplace{1cm}\fi
\caption[]{Non-\lte\ abundance corrections, lines included in
background opacities.
Solid curve: \bi\,250\,nm. Dashed curve: \bi\,209\,nm
}
\label{fig_abcorr_ag35}
\end{figure}

\begin{figure} 
\ifnum\doepsf=1\hspace*{.2cm}\epsfxsize=8.8cm\epsffile{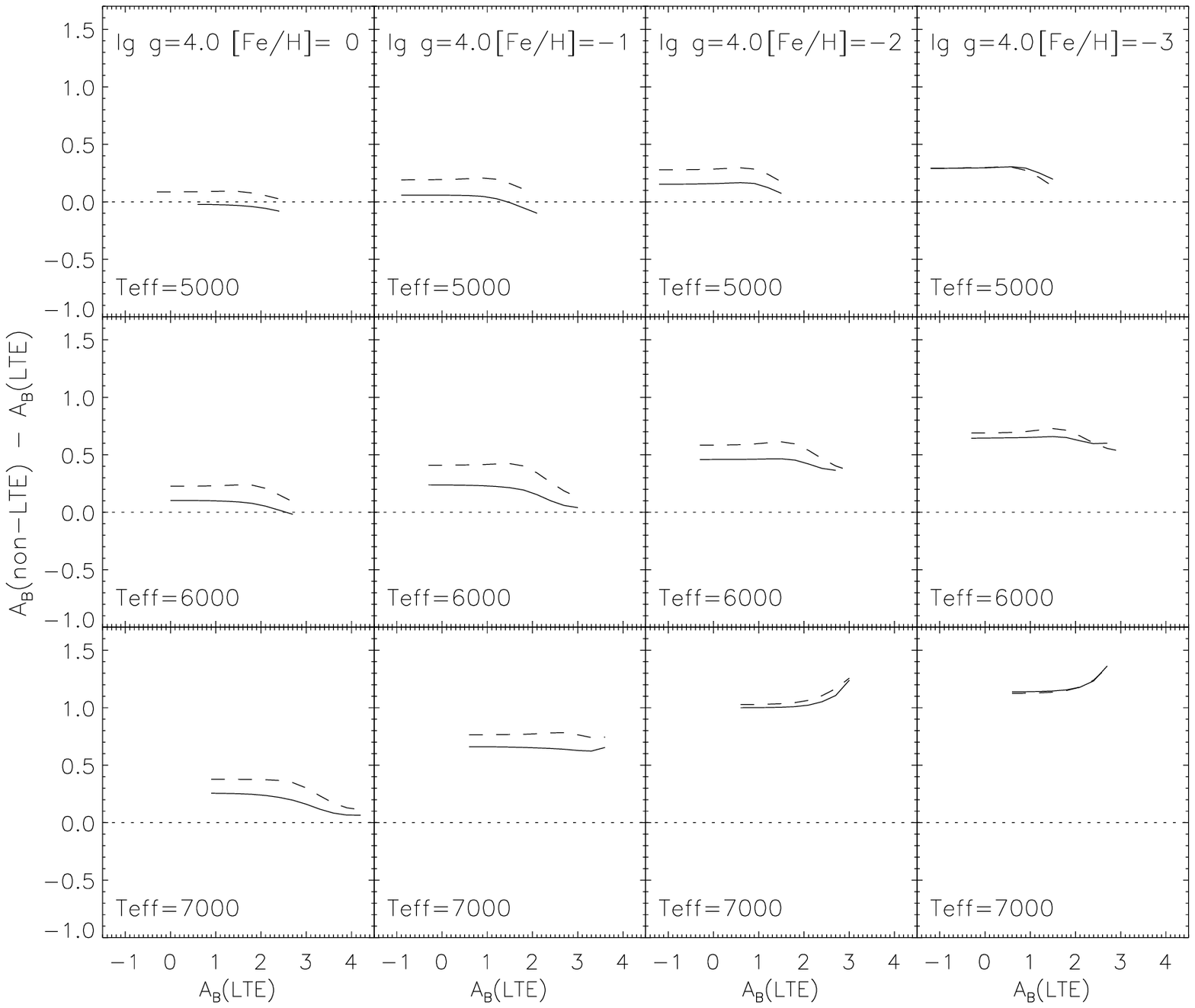}\else\picplace{1cm}\fi
\caption[]{Non-\lte\ abundance corrections, lines included in
background opacities.
Solid curve: \bi\,250\,nm. Dashed curve: \bi\,209\,nm
}
\label{fig_abcorr_ag4}
\end{figure}

\begin{figure}
\ifnum\doepsf=1\hspace*{.2cm}\epsfxsize=8.8cm\epsffile{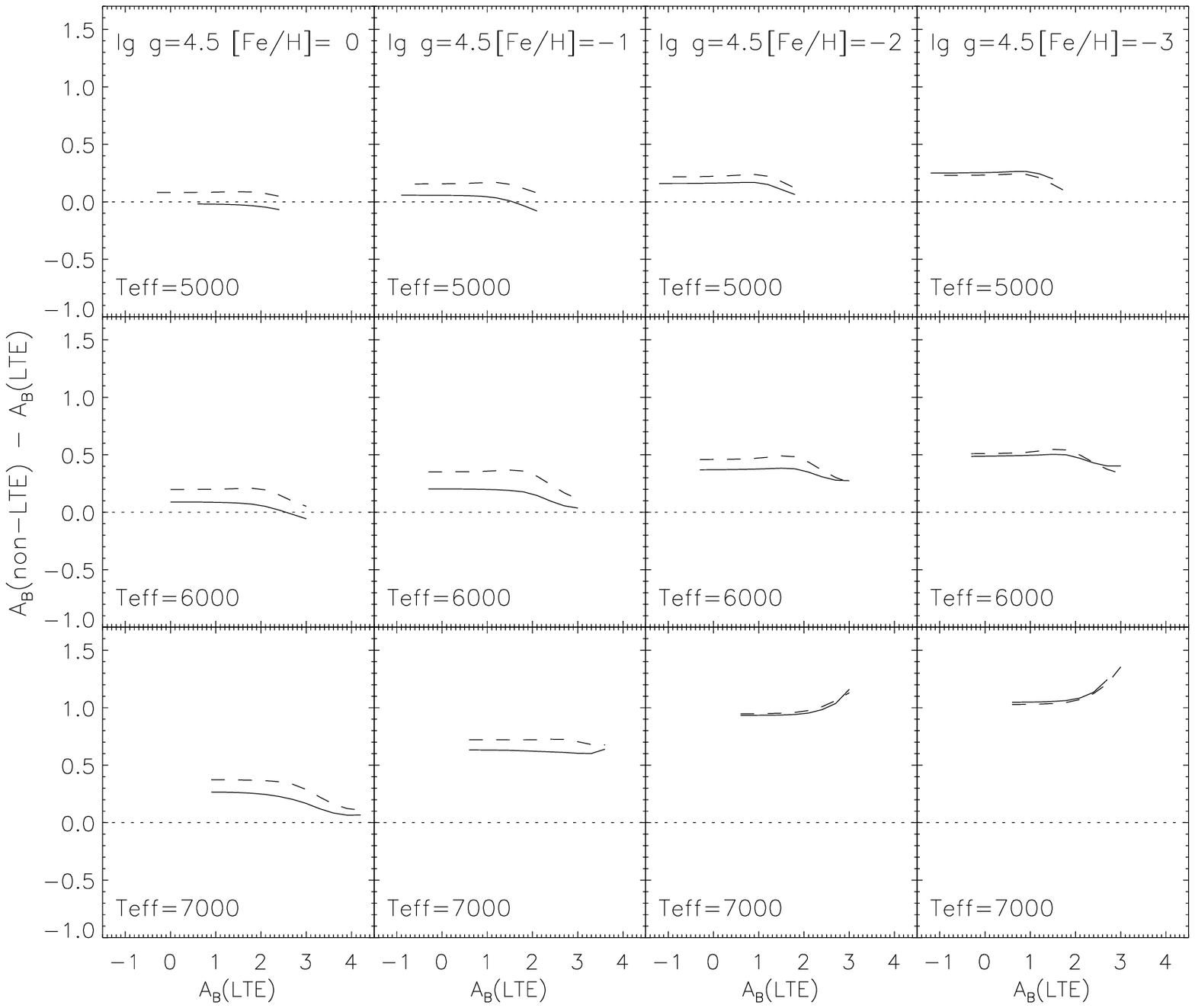}\else\picplace{1cm}\fi
\caption[]{Non-\lte\ abundance corrections, lines included in
background opacities.
Solid curve: \bi\,250\,nm. Dashed curve: \bi\,209\,nm
}
 \label{fig_abcorr_ag45}
\end{figure}

\begin{table}[tbh]
\caption[]{Computer routines and data files available via anonymous ftp}
\label{tab_files}
\begin{flushleft}
\begin{tabular}{ll}
\hline\noalign{\smallskip}
Internet adress: & ftp.astro.uio.no \\
Directory: & pub/boron \\
Data files: & adiffb0\_lte.dat, \\ 
            & adiffb1\_lte.dat \\
{\sc fortran} interpolation routine: & corrb\_nlte.f \\
{\sc idl} interpolation routine: & corrb\_nlte.pro \\
\noalign{\smallskip}
\hline
\end{tabular}
\end{flushleft}
\end{table}

\begin{figure}
\ifnum\doepsf=1\hspace*{.2cm}\epsfxsize=8.8cm\epsffile{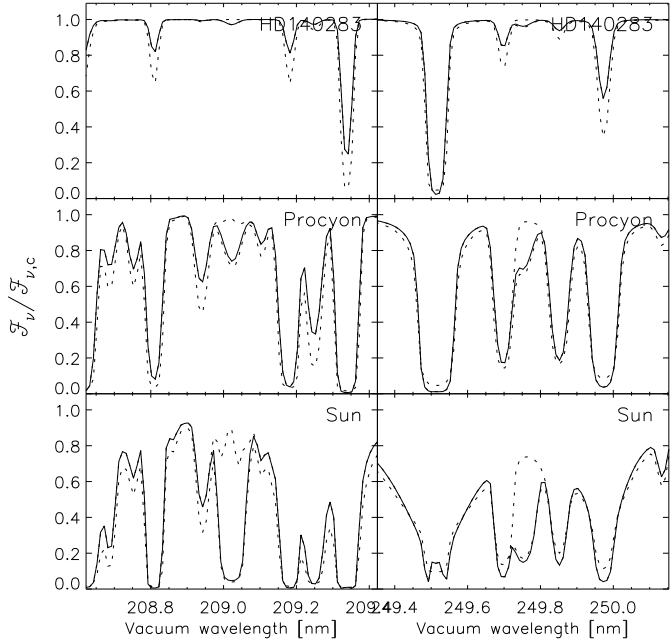}\else\picplace{1cm}\fi
\caption[]{Spectra for three stellar models computed with
the background lines in pure absorption (dotted curve) and with
a scattering part (solid curve). The boron abundances are
$-0.20$ (HD142083), $1.90$ (Procyon), $2.60$ (Sun), and, for the
scattering case also zero boron abundance (dotted curve). The normalisation
is the same as in Figs. \ref{fig_bspec_250} and \ref{fig_bspec_209}.
We stress again that what is showed here is not fits to observed
spectra but the results when our methods and line lists are applied to
our atmospheric models representing the three stars
}
 \label{fig_bspec_3}
\end{figure}
\begin{figure}
\ifnum\doepsf=1\hspace*{.2cm}\epsfxsize=13cm\epsffile{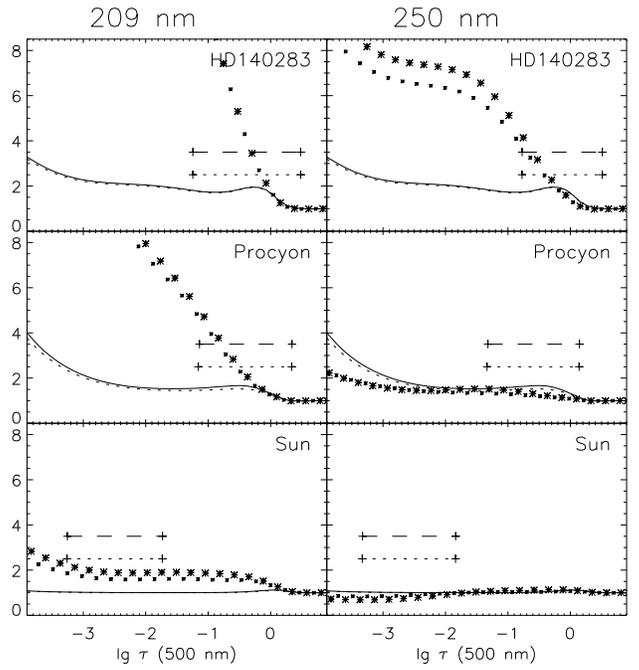}\else\picplace{1cm}\fi
\caption[]{Departure coefficients for three stellar models,
parameters as in Fig.~\ref{fig_bspec_3}.
The lines show $d_0 = N^*({\rm \bi})/N({\rm \bi})$, dotted for pure-absorption
background lines, solid for scattering contribution.
Stars show $S^l/B_\nu$ for the line indicated at top. 
Small stars: pure-absorption 
background lines. Big stars: scattering contribution. Crosses
show where the total optical depth in the line centre equals 1/3 and 3,
line styles identify background-line treatment as for $d_0$
}
\label{fig_depcoff_3}\end{figure}
\begin{figure}
\ifnum\doepsf=1\hspace*{.2cm}\epsfxsize=8.8cm\epsffile{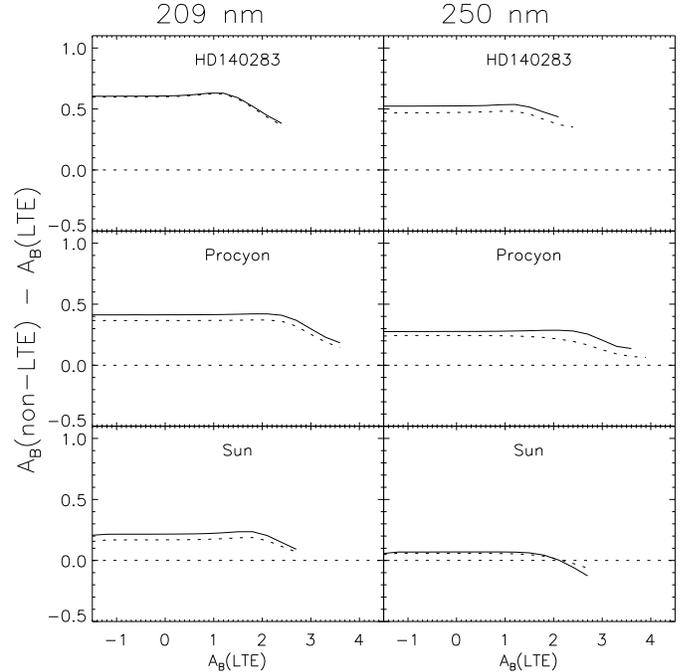}\else\picplace{1cm}\fi
\caption[]{Abundance corrections for the three stellar models.
Dotted curves for pure-absorption background lines, solid
curves for background with scattering
}
 \label{fig_abcorr_3}
\end{figure}

\begin{table}[tbh]
\caption[]{Non-\lte\ abundance correction for three specific
stellar models calculated with and without a scattering contribution in
the background-line opacities}
\label{tab_abcorr_3}
\begin{flushleft}
\begin{tabular}{lrrrrrr}
\hline\noalign{\smallskip}
Stellar model               & HD140283    & Procyon & Sun \\  
~~$T_{\rm eff}$ [K]  & 5680    & 6700    & 5780    \\
~~$\lg g$ [cgs]      & 3.50    & 4.03    & 4.44    \\   
~~[Fe/H]               & $-2.64$ & $-0.04$ & 0.00 \\ 
~~$\xi_t$ [km\,s$^{-1}$] & 1.5     & 2.5     & 1.15 \\
~~$A_{\rm B}$ \lte\     & $-0.20$ & 1.90    & 2.60 \\ 
\bi\,209\,nm \\
~~$\Delta A_{\rm B}$ no scattering & $+0.60$ & $+0.37$ & $+0.08$ \\  
~~$\Delta A_{\rm B}$ scattering    & $+0.61$ & $+0.42$ & $+0.11$ \\  
\bi\,250\,nm \\
~~$\Delta A_{\rm B}$  no scattering & $+0.47$ & $+0.22$ & $-0.05$ \\
~~$\Delta A_{\rm B}$ scattering    & $+0.52$ & $+0.29$ & $-0.10$ \\
\noalign{\smallskip}
\hline
\end{tabular}
\end{flushleft}
\end{table}

\section{Nature of LTE departures}
\subsection{Background}
The departures from \lte\ that Kiselman (1994) found for
the metal-poor atmosphere with low boron abundance
can be described as follows.

The source function of the resonance doublets is well described
by a two-level approximation, where radiative excitations and
deexcitations dominate over collisional processes. The hot, non-local,
ultraviolet radiation field causes an optical-pumping
effect where the excitation balance is shifted
upwards relative to \lte\ so that $S^l > B_\nu$.

The line opacities are set by the ionisation balance which departs
from the Saha equilibrium in that neutral boron is depleted --
overionisation. The pumping in the resonance lines causes
this because the radiation fields in the ionisation edges
of the excited levels are so much richer than at the ground-state
ionisation edge. Furthermore, these radiation fields are stronger
than the local Planckian value.

The pumping and the overionisation will both make absorption lines
weaker than in \lte, leading to underestimations of the boron
abundance when \lte\ is assumed.

\subsection{The line source function}
We calculate the two-level line source function

$$S^l_2 = (1-\varepsilon) \int \phi_\nu J_\nu
d\nu + \varepsilon B_\nu$$

where $$\varepsilon = {C_{\rm ul} \over C_{\rm ul} + A_{\rm ul}}$$ 
measures the destruction probability of line photons (stimulated
emission neglected).
The ratio $S^l/S^l_2$ was then studied as a measure of how good the
two-level approximation for each atmosphere and boron abundance
in the grid. 
It turned out that this ratio falls
very close to 1 for the largest part of the grid, both with and
without background lines, the exceptions are mainly for the cooler
stars in the case including background lines, where the ratio goes
down to $S^l/S^l_2 = 0.6$ for some models with low B abundance. This is because
the line source functions for the two doublet components are forced
to be equal by the strong coupling between their lower levels, while
$\int \phi_\nu J_\nu d\nu$ is quite different for the two components
 due to
the different strength of the background lines.
We consider the two-level picture adequate for discussing the
departures from \lte\ over the entire grid. 
This means that much
of the discussion of Kiselman (1994) is valid.

Over most of the grid, the \bi\ resonance lines experience pumping from
the hot, non-local, radiation field. The pumping effect decreases
as the boron lines grow stronger.

\subsection{Overionisation}
The relative overionisation at a given optical depth
increases towards higher temperatures, lower 
metallicities, and lower surface gravities. See
Figs.~\ref{fig_depcoff_n0g4} and \ref{fig_depcoff_n0g2},
where the overionisation is shown (solid curve) as the depletion factor
of neutral boron relative to the Saha number density: 
$d_0 = N^*$(\bi)$/N$(\bi). 
The increase towards
higher temperatures is largely due to the shift in the Saha 
equilibrium -- the relative impact of the overionisation
on the \bi\ abundance becomes
greater as boron becomes increasingly ionised.

When the \bi\ lines become strong, the pumping effect weakens leading
to less overionisation, as can
be seen when comparing Figs.~\ref{fig_depcoff_n0g4} and \ref{fig_depcoff_a0g4}.
The pumping-induced overionisation
 eventually turns into its opposite -- the ``photon suction'' 
described by Bruls et al. (1992) for Na\,{\sc i} and K\,{sc i} -- which
actually causes a marginal underionisation (overabundance
of neutral boron) in the coldest, 
metal-rich, atmospheres of high boron abundances. 

\subsection{Impact of background lines}
The introduction of background lines should decrease the optical
pumping effect simply by decreasing the local mean intensity
at the line frequency. This is also seen, compare
Fig.~\ref{fig_depcoff_n0g4} with Fig.~\ref{fig_depcoff_a0g4}.

\section{Non-LTE abundance corrections}

\subsection{Definitions}
We will here present non-\lte\ corrections that can be used to 
improve \lte\ abundance estimates. We give such corrections 
rather than, e.g., presenting non-\lte\ curves of growth because corrections
are less model-dependent, and because line blending makes
it difficult to unambiguously compute or measure equivalent widths needed
for a curve a growth.
The ordinary procedure to get such abundance corrections is to calculate
both \lte\ and non-\lte\ equivalent widths for a range of abundances.
The two resulting curves of growth are then used to interpolate the
\lte\
and the non-\lte\ abundances implied by a certain equivalent width. The
non-\lte\ abundance corrections can be tabulated as functions of \lte\
abundance or line strength and atmospheric parameters.

This procedure is straightforward when the background continuum of
the line that is studied varies little over the line width.
But when we introduce lines in the background opacities,
equivalent widths can be defined in several ways which will give
somewhat 
different results when curves of growth are compared. We choose here
to use a definition that essentially measures the area contained 
between the line and the background flux in a plot that is normalised
in the way of Figs.~\ref{fig_bspec_250} and \ref{fig_bspec_209}.
 We believe this to be the measure
that corresponds best with the result an abundance analyst
gets when fitting synthetic spectra to observations (Kiselman \&
Carlsson 1995).

In practice, we use the equivalent widths as they are
output from the program, their numerical values thus
including a dominating contribution from the background lines. 
For numerical reasons, we subtract the equivalent-width value from a run
with zero boron abundance. Care must also be taken to ensure that
the background spectrum is numerically identical in all results,
otherwise problems will arise when the boron lines are very weak.

\subsection{Results}
The plots of Figs.~\ref{fig_abcorr_ng4}-\ref{fig_abcorr_ag45}
display the non-\lte\ abundance corrections
as functions of \lte\ abundances for some cross sections of the grid. 
The range in \lte\ abundance for each plot is naturally limited by 
the boron abundance grid. We have also introduced limits
to ensure that the lines are strong enough to be of any observational
interest and
that they are weak enough for our modelling to be relevant.
The lower limit on line strength is $W < 0.1\,{\rm pm} = 1\,{\rm m{\AA}}$
(from the case without background lines). The upper limit is set to
assure that the line centre is optically thin at the uppermost 
point of the model atmosphere ($\lg \tau < -0.1$).

The curves in the plots are generally flat, illustrating that the
departures from \lte\ are set by the background radiation fields
as long as the \bi\ lines are weak. When the resonance lines become
strong enough to influence their own radiation fields, the 
pumping effects decreases, and so does the overionisation. 
This leads to the downturn in the abundance correction curves
seen in many of the plots.
In some cases, for the exotic parameter combination of hot, very
metal-poor stars with high boron abundance, there is a hint of
an upturn instead. The boron lines are here still rather weak
(cf. Figs.~\ref{fig_bspec_250} and \ref{fig_bspec_209}), and we
interpret the upturn as a result of the lines being formed
at greater heights in the atmosphere
where the relative overionisation is larger.

Figure~\ref{fig_abcorr_ng4} displays the corrections without 
background lines for $\lg g =
4.0$. Compare this with Fig.~\ref{fig_abcorr_ag35} to see the
importance of the background lines.

The corrections calculated for the entire grid are available,
together with interpolation routines, via ftp from 
ftp.astro.uio.no, directory pub/boron. 
The routines and formats
are similar to those presented by Carlsson et al. (1994)
for Li abundance corrections. More details are given in Table~\ref{tab_files}.

For quick reference, we also present numerical values for the 
corrections (\bi\,250\,nm) over a part of the
grid representing solar-like stars in Table~\ref{tab_tab}.

\subsection{Uncertainties}

\subsubsection{General}

The model-atom uncertainties are discussed by Kiselman (1994)
who concludes that they are not likely to be the most important
cause of errors.

Our corrections are model dependent and should really
only be applied to \lte\ results acquired with the same atmospheric
models.
Model-atmosphere errors and differences, within the plane-parallel paradigm,
 will, however, probably influence the non-\lte\
corrections only to second order.
Effects caused by granulation and other kinds
of photospheric inhomogeneity is probably a matter of greater concern.
(And would be so even if the lines were formed in perfect \lte.)

The 250\,nm doublet is approaching optical thickness in the uppermost point
of our solar-like atmospheric models with high boron abundance
when background lines are included. The modelling is then not very
realistic, especially since the lines should be affected by 
a chromospheric temperature rise. Everything seems to imply, 
however, that departures from \lte\ will be small for 
stars similar to the Sun.

\subsubsection{Background lines}
\label{sec_back_lines}
Errors in the line lists and inadequacies in our treatment of
background lines influence the results in two ways.
First, errors in oscillator strengths and non-\lte\ effects can cause
the background opacities to be erroneous. This could in principle
be mended by detailed fitting to observed spectra. 
(Something which would be
needed anyhow to allow for the detailed abundance pattern of
individual stars.)
Second, errors in the background source function are introduced
by treating the line opacities as pure absorption. This
will tend to thermalise the transitions involved.
The effect of photon escape, which would lead to $S^l < B_\nu$
for strong lines is expected to be suppressed leading to
a line source function
of the boron line that is too close to the \lte\ value. 
Abundance corrections will then be close to zero instead of
negative (if overionisation is negligible). 

We have investigated the latter effect by introducing a scattering
part in the background line opacities via a line-photon
destruction probability $\varepsilon = {C_{\rm ul} \over C_{\rm ul} +
A_{\rm ul}}$ that is
calculated using Van Regemorter's (1962) approximation for
the collisional rate. Thus the absorption fraction of the
line opacity is
$$ {\kappa_l \over \kappa_l + \sigma_l} = \varepsilon$$
and the scattering fraction
$$ {\sigma_l \over \kappa_l + \sigma_l} = 1 - \varepsilon.$$

This was done for atmospheric models representing the Sun,
Procyon and the halo star HD140283 (the same models as in 
Kiselman 1994) and the same range of B abundances as in the
main grid. The results are illustrated in Figs.~\ref{fig_bspec_3}, 
\ref{fig_depcoff_3}, \ref{fig_abcorr_3}, and Table~\ref{tab_abcorr_3}.

Figure~\ref{fig_bspec_3} compares the appearance of 
spectra as computed with (solid curve) and without 
(dotted curve) scattering in background lines.

The relevant departure coefficients are shown in Fig.~\ref{fig_depcoff_3}
in the same way as in
Figs.~\ref{fig_depcoff_n0g4}--\ref{fig_depcoff_a0g4}. 
The most significant difference is in the line source
function of the 250\,nm line.

Figure~\ref{fig_abcorr_3} displays non-\lte\ abundance 
corrections as function
of derived \lte\ abundance (in a similar way as 
Figs.~\ref{fig_abcorr_ng4}--\ref{fig_abcorr_ag45}) for
the cases with (solid curves) and without (dashed curves)
scattering in background lines.
The effect of introducing the background scattering is to
increase the magnitude of the abundance corrections, regardless
whether these are positive or negative. (Note the crossing of
the curves at the zero level for the solar 250\,nm case.)
This is expected since we decrease the photon-destruction
probability and thus increase the effective photon path length
-- the non-local nature of the radiation field is enhanced.
The differences in abundance corrections
are, however, small -- never more than 0.1\,dex

We conclude that a refined treatment of background lines
may make the non-\lte\ corrections slightly greater. We note, however,
that any such refinement would need adjustments of the line lists
in order for the treatment to be consistent.

\begin{table}[tbh]
\caption[]{Reanalysis of literature data. The sources for
the \lte\ boron abundances and atmospheric parameters 
are: $a$) Edvardsson et al. 1994, 
$b$) Duncan et al. 1992, $c$) Lemke et al. 1993, $d$)
Kohl et al. 1977. The non-\lte\ abundances in column N were calculated
without background lines. In column A background lines were included
(these are our recommended abundances). The A value for the Sun is
given although the line core is optically thick at the uppermost
atmospheric point at this abundance }
\label{tab_reana}
\begin{flushleft}
\begin{tabular}{llrrrrrr}
\hline\noalign{\smallskip}
Star &   $T_{\rm eff}$ & $\lg g$ & [Fe/H] & $A_{\rm B}$
& $A_{\rm B}$ & $A_{\rm B}$ \\
     & [K]           & [cgs]   & & \lte\ & N~ & A~ \\
\noalign{\smallskip}
\hline\noalign{\smallskip}
HD140283$^a$     & 5680 & 3.50 & $-2.64$ & $-0.20$ & 0.39 & $0.29$ \\
HD19445$^b$      & 5880 & 4.40 & $-2.15$ & $0.40$  & 0.87 & $0.76$ \\
HD201891$^b$     & 5870 & 4.50 & $-1.06$ & $1.70$  & 2.06 & $1.86$ \\
$\theta $UMa$^c$ & 6380 & 4.09 & $-0.14$ & $2.30$  & 2.59 & $2.43$ \\
Procyon$^c$      & 6700 & 4.03 & $-0.04$ & $1.90$  & 2.28 & $2.10$ \\
$\iota $Peg$^c$  & 6750 & 4.35 & $-0.04$ & $2.40$  & 2.77 & $2.58$ \\
Sun$^d$          & 5780 & 4.44 &   0.00  & 2.60    & 2.64 & $2.55$ \\
\noalign{\smallskip}
\hline
\end{tabular}
\end{flushleft}
\end{table}
\begin{figure}[tbh]
\ifnum\doepsf=1\hspace*{.2cm}\epsfxsize=8.8cm\epsffile{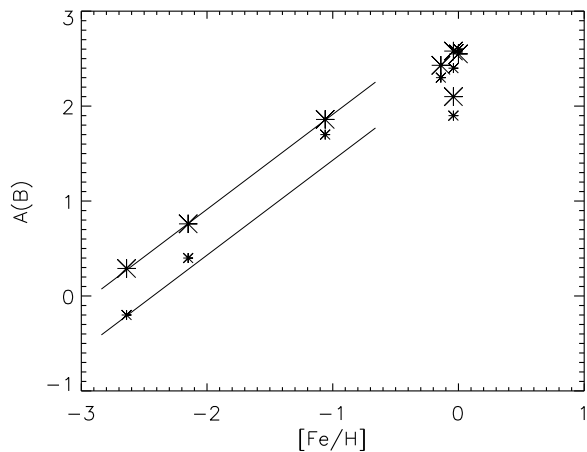}\else\picplace{1cm}\fi
\caption[]{Boron abundances for solar-type stars plotted as a
function of metallicity. Data from
Table~\ref{tab_reana}, small stars are \lte\ abundances. Two lines
with B/Fe $=$ constant are drawn for comparison
}
\label{fig_bstars}
\end{figure}

\begin{table}[h]
\caption[]{Numerical non-\lte\ abundance corrections (in dex) for
\bi\ 250\,nm for part of the atmosphere grid (computed 
with background lines). The corrections should be added to
\lte\ abundances. Gaps in the tables mean that
the line is too weak or too strong according to the 
criteria described in section 4.2, or simply that the (LTE)
abundance falls outside the abundance range   }
\label{tab_tab}
\begin{flushleft}
\begin{tabular}{lrrrrrrrr}
\hline\noalign{\smallskip}
\noalign{\ ~~~~~~~~~~~~~~~~~~~~~~~~~~\ $A_{\rm B}({\rm LTE})$}
$T_{\rm eff}/\lg g/$[Fe/H] & 0.00 & 0.90 & 1.50 & 2.10 & 2.40 \\
\ [K]/[cgs]/ \\
\noalign{\smallskip}
5000/2.00/~~$0.00$ &$-$0.01&$-$0.02&$-$0.03&       &        \\
5000/2.00/$-1.00$ &$+$0.10&$+$0.02&$-$0.06&       &        \\
5000/2.00/$-2.00$ &$+$0.23&$+$0.09&$-$0.03&       &        \\
5000/2.00/$-3.00$ &$+$0.49&$+$0.32&$+$0.15&       &        \\
5000/3.50/~~$0.00$ &       &$-$0.03&$-$0.04&$-$0.07&        \\
5000/3.50/$-1.00$ &$+$0.07&$+$0.05&$-$0.02&       &        \\
5000/3.50/$-2.00$ &$+$0.17&$+$0.15&$+$0.04&       &        \\
5000/3.50/$-3.00$ &$+$0.34&$+$0.31&$+$0.19&       &        \\
5000/4.50/~~$0.00$ &       &$-$0.02&$-$0.03&$-$0.05&$-$0.07 \\
5000/4.50/$-1.00$ &$+$0.06&$+$0.05&$+$0.01&$-$0.08&        \\
5000/4.50/$-2.00$ &$+$0.16&$+$0.17&$+$0.11&       &        \\
5000/4.50/$-3.00$ &$+$0.25&$+$0.26&$+$0.20&       &        \\
\noalign{\smallskip}
6000/2.00/~~$0.00$ &$+$0.12&$+$0.11&$+$0.09&$+$0.05&$+$0.02 \\
6000/2.00/$-1.00$ &$+$0.27&$+$0.25&$+$0.21&$+$0.13&$+$0.08 \\
6000/2.00/$-2.00$ &$+$0.55&$+$0.54&$+$0.51&$+$0.43&$+$0.38 \\
6000/2.00/$-3.00$ &$+$0.96&$+$0.96&$+$0.96&$+$0.93&$+$0.97 \\
6000/3.50/~~$0.00$ &$+$0.11&$+$0.11&$+$0.10&$+$0.05&$+$0.02 \\
6000/3.50/$-1.00$ &$+$0.26&$+$0.25&$+$0.23&$+$0.16&$+$0.10 \\
6000/3.50/$-2.00$ &$+$0.54&$+$0.54&$+$0.54&$+$0.49&$+$0.45 \\
6000/3.50/$-3.00$ &$+$0.81&$+$0.81&$+$0.82&$+$0.79&$+$0.79 \\
6000/4.50/~~$0.00$ &$+$0.09&$+$0.09&$+$0.08&$+$0.05&$+$0.02 \\
6000/4.50/$-1.00$ &$+$0.20&$+$0.20&$+$0.19&$+$0.15&$+$0.10 \\
6000/4.50/$-2.00$ &$+$0.37&$+$0.37&$+$0.38&$+$0.35&$+$0.31 \\
6000/4.50/$-3.00$ &$+$0.49&$+$0.49&$+$0.50&$+$0.47&$+$0.43 \\
\noalign{\smallskip}
7000/2.00/~~$0.00$ &       &$+$0.20&$+$0.19&$+$0.18&$+$0.17 \\
7000/2.00/$-1.00$ &       &$+$0.61&$+$0.61&$+$0.60&$+$0.59 \\
7000/2.00/$-2.00$ &       &$+$1.11&$+$1.11&$+$1.12&$+$1.13 \\
7000/2.00/$-3.00$ &       &$+$1.35&$+$1.36&$+$1.37&$+$1.39 \\
7000/3.50/~~$0.00$ &       &$+$0.24&$+$0.24&$+$0.22&$+$0.21 \\
7000/3.50/$-1.00$ &       &$+$0.67&$+$0.67&$+$0.66&$+$0.65 \\
7000/3.50/$-2.00$ &       &$+$1.05&$+$1.05&$+$1.07&$+$1.09 \\
7000/3.50/$-3.00$ &       &$+$1.21&$+$1.22&$+$1.25&$+$1.30 \\
7000/4.50/~~$0.00$ &       &$+$0.27&$+$0.26&$+$0.25&$+$0.23 \\
7000/4.50/$-1.00$ &       &$+$0.63&$+$0.63&$+$0.62&$+$0.62 \\
7000/4.50/$-2.00$ &       &$+$0.93&$+$0.94&$+$0.95&$+$0.98 \\
7000/4.50/$-3.00$ &       &$+$1.05&$+$1.05&$+$1.08&$+$1.13 \\
\noalign{\smallskip}
\hline
\end{tabular}
\end{flushleft}
\end{table}

\section{Impact on existing observations}
Table~\ref{tab_reana} features \lte\ boron abundances from the literature
and the abundances after the non-\lte\ corrections
from this paper have been applied.
The non-\lte\ corrections have been interpolated
from the grid, which explains the small differences for
the three stars of Table~\ref{tab_abcorr_3}.
The abundance data is displayed in Fig.~\ref{fig_bstars} as a plot of boron 
abundance against metallicity for the stars of Table~\ref{tab_reana}.

The inclusion of background lines moderates somewhat
the magnitude of the non-\lte\ effects for HD140283
and Procyon found by Kiselman (1994).
The change in abundance for the extreme
halo star HD140283 is small, however, and the conclusions
of Edvardsson et al. (1994) regarding the boron/beryllium
ratio are unaffected.
For Procyon, it seems that the new results reconfirm the
general conclusion of Lemke et al. (1993) that this star really
is somewhat boron depleted relative to the Sun.

The new result for the Sun indicates a slightly lower boron
abundance than before. Note, however, that we have adopted the rather
uncertain photospheric
\lte\ abundance of $2.60\pm 0.3$ that was derived by Kohl et
al. (1978). The meteoritic abundance is $2.88\pm 0.04$ 
(Anders \& Grevesse 1989) and may perhaps be more realistic.

It is interesting to note that, for the three metal-poor stars, 
the non-\lte\ corrected [B/Fe] is close to constant,
in contrast to the \lte\ results, see Fig.~\ref{fig_bstars}.

\section{Conclusions}
We have presented a study of departures from \lte\ for
neutral boron in a grid of cool stellar atmospheric models.
The processes responsible for these departures over the grid are
essentially the same as those described by Kiselman (1994).
The non-\lte\ abundance corrections are generally
greatest for hot, metal-poor,
and low-gravity stars, i.e. they become more important as
the \bi\ lines get weaker for a fixed B abundance.
\lte\ abundances based on \bi\ lines should always be corrected
for these effects. We have presented such corrections
for cool stars and made them available in computer-readable
form.

\begin{acknowledgements}
We thank Bengt Edvardsson for help with line lists and 
valuable comments on the manuscript. Thanks are also
due to P{\aa}l Brekke for providing the solar UV spectrum.
\end{acknowledgements}


\begin{thebibliography}{}
\bibitem[1989]{ag} Anders E., Grevesse N., 1989,
 Geochim. Cosmochim. Acta, 53, 197
\bibitem[1994]{b} Bell R.A., Paltoglou G., Tripicco M.J., 1994,
 MNRAS, 268, 771
\bibitem[1995]{b} Bell R.A., Tripicco M.J., Synthetic Spectra,
Synthetic colours and line lists. In: Adelman S.J., Wiese W.L. (eds.),
Astrophysical Applications of Powerful New Databases, ASP Conference
Series, Vol. 78, p.367
\bibitem[1991]{b} Bi\'emont E., Baudoux M., Kurucz R.L., Ansbacher W.,
 Pinnington E.H., 1991, A\&A, 249, 539
\bibitem[1992]{b} Bruls J.H.M.J., Rutten R.J., Shchukina N.G., 1992,
A\&A, 265, 237
\bibitem[1992]{bk} Buser R., Kurucz R.L., 1992, A\&A, 264, 557
\bibitem[1986]{mc} Carlsson M., 1986, Uppsala Astronomical
 Observatory Report No. 33 
\bibitem[1994]{crbs} Carlsson M., Rutten R.J., Bruls J.H.M.J.,
 Shchukina N.G., 1994, A\&A, 288, 860
\bibitem[1993]{dll} Duncan D.K., Lambert D.L., Lemke M., 1992, ApJ,
 401, 584
\bibitem[1995]{de} Duncan D.K., Peterson R.C., Thorburn J.A.,
 Pinsonneault M.H., Deliyannis C.P., 1995, Boron in the Hyades giants.
 In: Crane P. (ed.), The Light Elements Abundances, Springer, p. 425
\bibitem[1993]{e} Edvardsson B., Andersen J., Gustafsson B., et al., 1993, A\&A, 275, 101
\bibitem[1994]{e} Edvardsson B., Gustafsson B., Johansson S.G., et
al., 1994, A\&A, 290, 176
\bibitem[1975]{g} Gustafsson B., Bell R.A., Eriksson K.,
 Nordlund {\AA}., 1975, A\&A, 42, 407
\bibitem[1995]{g} Gustafsson B., 1995, Opacity incompleteness and
atmospheres of late-type stars. In: Adelman S.J., Wiese W.L. (eds.),
Astrophysical Applications of Powerful New Databases, ASP Conference
Series, Vol. 78, p.347
\bibitem[1990]{h} Holweger H., Heise C., Kock M., 1990, A\&A, 232, 510
\bibitem[1993]{j} Johansson S.G., Litz\'en U., Kasten J., Kock M.,
 1993, ApJ, 403, L25
\bibitem[1977]{k} Kohl J.L., Parkinson W.H., Withbroe G.L., 1977, ApJ,
 212, L101
\bibitem[1994]{dk} Kiselman D., 1994, A\&A, 286, 169
\bibitem[1995]{dm} Kiselman D., Carlsson M., 1995, Non-LTE effects on Be and
B abundance determinations in cool stars. In: Crane P. (ed.),
The Light Elements Abundances, Springer, p.372
\bibitem[1991]{k} Kurucz R.L., 1991, New opacity calculations. In:
Crivellari L. (ed.), Stellar atmospheres: Beyond Classical Models,
Kluwer, p.441
\bibitem[1993]{lle} Lemke M., Lambert D.L., Edvardsson B., 1993, PASP,
 105, 468
\bibitem[1986]{oa} Olson G.L., Auer L.H., Buchler J.R.,1986, JQSRT, 35, 431
\bibitem[1985]{sc} Scharmer G.B., Carlsson, M., 1985, J.Comput.Phys. 59, 56 
\bibitem[1962]{vr} Van Regemorter H., 1962, ApJ, 136, 906
\end{thebibliography}
\end{document}